\documentclass[aps,pra,showpacs,twocolumn]{revtex4}
\usepackage{graphics}
\usepackage{amssymb}
\usepackage{amsmath}
\usepackage{bm}

\usepackage[pdftex]{graphicx}

\begin{document}

\title[]{Free Expansion of Quasi-2D Bose-Einstein Condensates with Quantized Vortices}

\author{Sang Won Seo, Jae-yoon Choi, and Yong-il Shin}\email{yishin@snu.ac.kr}

\affiliation{Center for Subwavelength Optics and Department of Physics and Astronomy, Seoul National University, Seoul 151-747, Korea}

\date{\today}

\begin{abstract}
We observe that a density-depleted vortex core evolves into concentric density ripples in a freely expanding quasi-two-dimensional Bose-Einstein condensate. The atomic density of the expanding condensate rapidly reduces due to the fast expansion along the tightly confining axial direction, and thus the transverse expansion of the condensate is not significantly affected by the atom-atom interactions. We find that the observed density profiles of the vortex cores are in good quantitative agreement with numerical simulations assuming no interactions. We analyze the defocused images of the vortex cores, where defocussing is caused by the free fall of the condensate during the expansion time. In the defocused images, the vortex core becomes magnified and appears to be filled after a certain expansion time which depends on the vortex charge number.
\end{abstract}


\maketitle

\section{Introduction}

Quantized vortices are topological defects in a superfluid, possessing quantized circulation of the superfluid around them. They play an important role in the transport phenomena in the superfluid such as the critical superflow and quantum turbulence~\cite{RJD}. Furthermore, they are responsible for the superfluid phase transition in two dimensions, where pairing of vortices with opposite circulation occurs below a critical temperature~\cite{BKT}. In ultracold atom experiments, quantized vortices and related phenomena have been actively studied~\cite{BPAresource}. They were first observed in rotating Bose-Einstein condensates (BECs)~\cite{Dalibardvortex,Jamil} and many aspects of their dynamic behavior have been investigated, including line defect deformation~\cite{linedeformation}, instability of multiply charged vortices~\cite{vortexsplit,quadcharge}, and vortex dipole dynamics~\cite{vortexdipole,realtime}. Recently, thermally activated vortices have been observed in two-dimensional (2D) Bose gases~\cite{Pointdefect,BKTpaper}.

At the core of a quantized vortex, superfluid density is depleted because of the diverging centrifugal potential. This distinctive feature is typically used for detecting the quantized vortices. In a trapped BEC, however, the spatial extent of the vortex core, which is determined by the healing length of the condensate, is generally smaller than the wavelength of imaging light, and so its {\it in situ} optical detection is experimentally challenging. A conventional method to circumvent this difficulty is releasing the trap to let the condensate freely expand, where the vortex core also expands. Free expansion of a BEC containing a quantized vortex was theoretically studied, showing that the vortex core may expand faster than the condensate during the initial stage of the expansion~\cite{LPS98}, and further faster in the case of an oblate BEC~\cite{DM00}.

In this paper, we study expansion dynamics of a quasi-2D, {\it i.e.} highly oblate BEC containing quantized vortices. Because the condensate expands fast along the tight confinement direction at the beginning of the free expansion, the atomic density and consequently the atom-atom interaction effects rapidly reduce in the expanding condensate. For this sudden quenching of the interactions, the density-depleted vortex core, which is maintained to be small in the trapped condensate by the interactions, will evolve differently from that in three-dimensional hydrodynamic expansion. We observe that the ring-shaped density ripples develop surrounding the vortex core in a freely expanding quasi-2D condensate and find that their time evolution is quantitatively well described by numerical simulations assuming no interactions. Finally, we investigate the defocus effect in imaging of the vortex core, which is caused by the free fall of the condensate during the expansion time.

\section{Experiment}

We prepare a quasi-2D Bose-Einstein condensate containing quantized vortices as described in our previous publication~\cite{PhaseFluc}. Bose-Einstein condensates of $^{23}$Na atoms in the $|F=1,m_F=-1\rangle$ state are generated in an optically plugged magnetic quadrupole trap~\cite{SNUBEC}, and transferred into a single optical dipole trap whose trapping frequencies are $(\omega_x, \omega_y, \omega_z) = 2\pi \times$(3.0, 3.9, 370)~Hz, where the $z$ direction is along the gravity. For a typical atom number of $1.0\times 10^6$, the chemical potential is $\mu\approx h\times 260$~Hz in the Thomas-Fermi (TF) approximation. This is less than the confining energy $\hbar \omega_z$, and at low temperatures our sample constitutes a weakly interacting quasi-2D BEC. The healing length is $\xi=\hbar/\sqrt{2m\mu} \approx 1~\mu$m ($m$ is the atomic mass), which is the characteristic spatial extent of a density-depleted vortex core in the condensate.

In order to induce vortex excitations, we intentionally displace the optical trap from the center of the plugged magnetic trap by about $40~\mu$m in the horizontal direction, and carry out a rapid transfer of the condensate to the optical trap. The transfer is made by ramping up the optical trap and simultaneously ramping down the magnetic potential for 500~ms. During this transfer, the optical plug beam is kept on to let the condensate experience an obstacle as it flows into the optical trap. Finally, the plug beam is linearly turned off for 400~ms. Free expansion is initiated by suddenly turning off the optical trap, and the column density distribution is measured by taking an absorption image along the $z$ axis after an expansion time $t_e$.

\begin{figure}
\includegraphics[width=5.6cm]{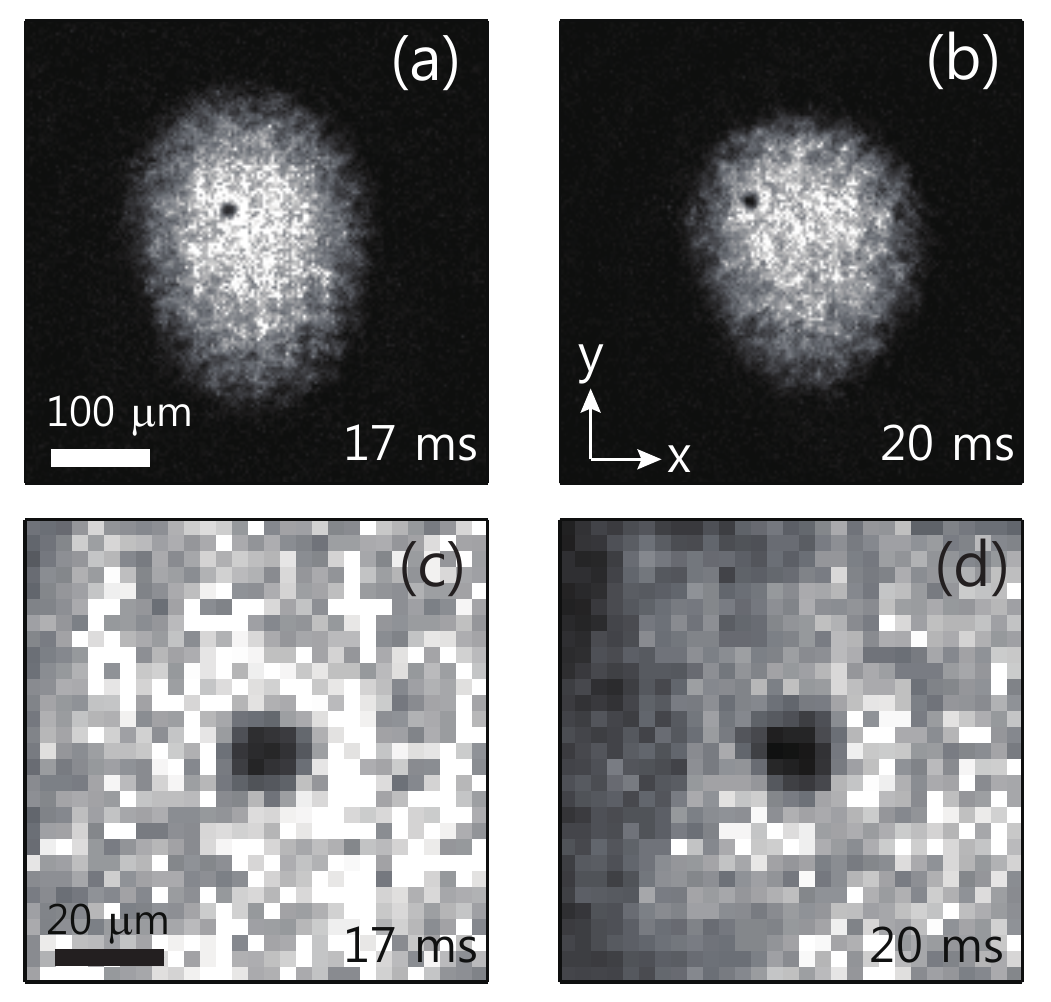}
\caption{
Optical density images of quasi-2D Bose-Einstein condensates after a time-of-flight of (a) $t_e=$17~ms and (b) 20~ms. The condensates are mechanically perturbed and prepared in nonequilibrium states (see text for detail), and density-depleted vortex cores are observed in the freely expanding condensates. Images of the vortex core for (c) $t_e=$17~ms and (d) 20~ms, obtained by averaging several images like (a) and (b), respectively.}
\label{fig1}
\end{figure}

Right after being transferred into the optical trap, the condensate showed very complicated and random density modulations in expansion, indicating that it was strongly perturbed in the transfer process~\cite{PhaseFluc}. We let this non-equilibrium state relax by holding the condensate in the optical trap. After long relaxation times over 15~s, density-depleted vortex cores were observed clearly in the expanding condensate (Fig.~\ref{fig1}). At this moment, the temperature of the thermal component was about 50~nK and the core visibility in the center region of the condensate was above 70\%. We note that such clear vortex cores were not observed when condensates were prepared in thermal equilibrium at this temperature, which is achieved by transferring thermal atoms into the optical trap and applying slow evaporative cooling.

The density distributions of the vortex core for various expansion times are obtained by averaging several optical density images of the vortex cores located in the center region of the condensates [Fig.~1(c) and (d)]. It is observed that a noticeable density hump forms around the core and propagates outward in the course of free expansion.

\begin{figure}
\includegraphics[width=8cm]{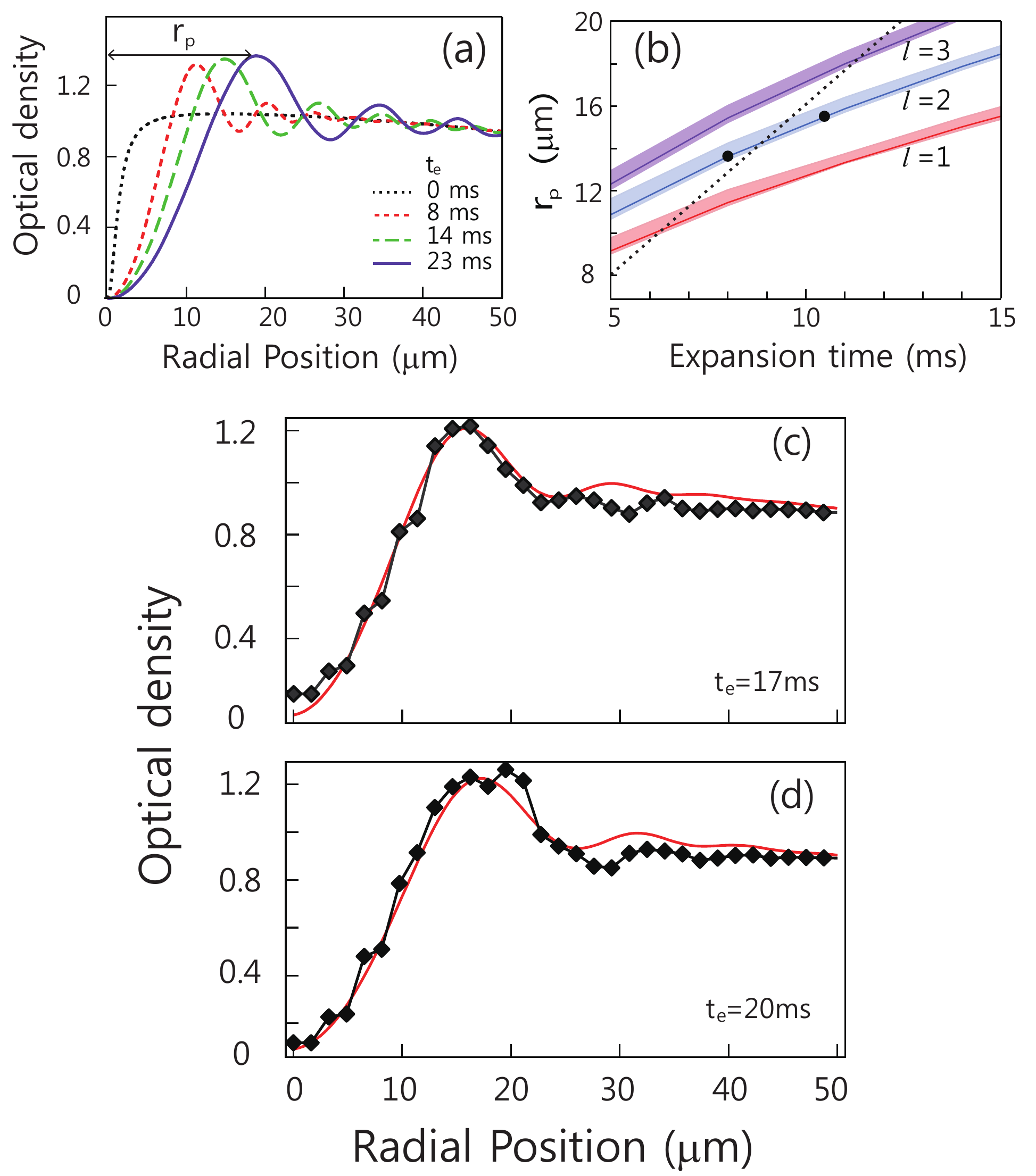}
\caption{
(Color online) (a) Numerical results for the radial profiles of a freely expanding, non-interacting condensate containing a singly-charged ($l$=1) vortex for various expansion times $t_e$, where $\xi=1~\mu$m. (b) Ring radius $r_p$ versus expansion time $t_e$ for various phase winding numbers $l$. The solid lines are for $\xi=1~\mu$m and the shaded areas correspond to $\xi=0.5 \sim 2~\mu$m. The dotted line denotes the critical radius to have a filled core in the defocused image for $d=\frac{1}{2}gt_e^2$ in Fig.~3(a). (c,d) Radial profiles (black diamond) obtained by azimuthally averaging the images in Fig.~1(c) and (d), respectively. The red lines are the numerical results taking into account the finite spatial resolution of the imaging system.
}
\label{fig2}
\end{figure}

\section{Evolution of the Vortex Core}

To understand the observed evolution of the vortex core, we perform numerical simulation of free expansion of a BEC containing a quantized vortex at its center. Here, in particular, we assume that the atom-atom interactions immediately vanish at the start of the expansion. In this non-interacting case, the axial motion in the $z$ direction is separable from the transverse motion in the $x$-$y$ plane, so we consider only the transverse expansion of the condensate. The initial condensate wave function is set as
\begin{equation}\label{eq1}
\psi_0{(r,\phi)}=\frac{r~e^{il\phi}}{\sqrt{2 l^2 \xi^2  +r^2}}\times \sqrt{ n_0 \max [1-\frac{r^2}{r_\textrm{TF}^2},0]},
\end{equation}
where the first part describes the vortex core structure with phase winding number $l$~\cite{Pethick} and the second part corresponds to the TF density profile in a harmonic potential with central density $n_0$ and radius $r_\textrm{TF}$. Because of the cylindrical symmetry of the condensate wave function, its non-interacting evolution can be expressed as
\begin{equation}\label{eq2}
\psi{(r,\phi,t_e)}=e^{il\phi}\sum_{n=1}^{\infty}c_{nl}J_l{(\frac{\beta_{nl}}{R} r)}\exp{\left(-i\frac{\hbar}{2m} \frac{\beta_{nl}^2}{R^2} t_e\right)},
\end{equation}
where $J_l$ is the Bessel function of the first kind of order $l$, and  $\beta_{nl}$ is the $n$th zero of $J_l$. The coefficients $c_{nl}$ are determined by $c_{nl}=\frac{2e^{-il\phi}}{R^2\left [J_{l+1}{(\beta_{nl})} \right ]^2}\int_{0}^{R}\psi_0(r) J_l{(\beta_{nl}\frac{r}{R})}rdr$. In our calculation, we set $r_\textrm{TF}=150~\mu$m and $R=3r_\textrm{TF}$.

Figure~\ref{fig2}(a) shows the results of the numerical calculation for the density distribution $n(r,t_e)=|\psi(r,t_e)|^2$ with $\xi=1~\mu$m and $l=1$. A multiple-ring structure develops and radially expands as observed in the experiment. Furthermore, we find that the radial profiles obtained by azimuthally averaging the vortex core images in Fig.~\ref{fig1}(c) and (d) are in good quantitative agreement with the numerical results taking into account the finite spatial resolution of 5~$\mu$m in our imaging system [Fig.~\ref{fig2}(b) and (c)]. This demonstrates that the interaction effects are negligible in the evolution of the vortex core in the freely expanding quasi-2D Bose-Einstein condensate.

Next, we numerically investigate the free expansion of a quasi-2D condensate containing multiply charged vortices with $l>1$. A concentric, multiple-ring structure develops as in the $l=1$ case, but its expansion rate increases with higher $l$. We characterize the time evolution of the concentric ripples with the radius $r_p$ of their first density-peak ring [Fig.~2(a)], and find that the ring expansion is nicely fit to $r_p=(3.9, 4.7, 5.4)\times \sqrt{t_e}~\mu$m/ms$^{1/2}$ for $l=1, 2, 3$, respectively. The $\sqrt{t_e}$-dependence can be understood from the continuity equation for 2D propagation, requiring $r_p \dot{r_p}=\textrm{constant}$. The dependence of the expansion rate $\dot{r_p}$ on the initial core size, {\it i.e.} $\xi$ is very marginal. This seems to reflect the weak dependence of the kinetic energy of the vortex state on the core size, $E_v \propto l^{2} \ln (r_\textrm{tf}/\xi)$ and suggests that the vortex charge number $l$ can be unambiguously determined from the radius $r_p$ for a given expansion time $t_e$.

\begin{figure}
\includegraphics[width=8.8cm]{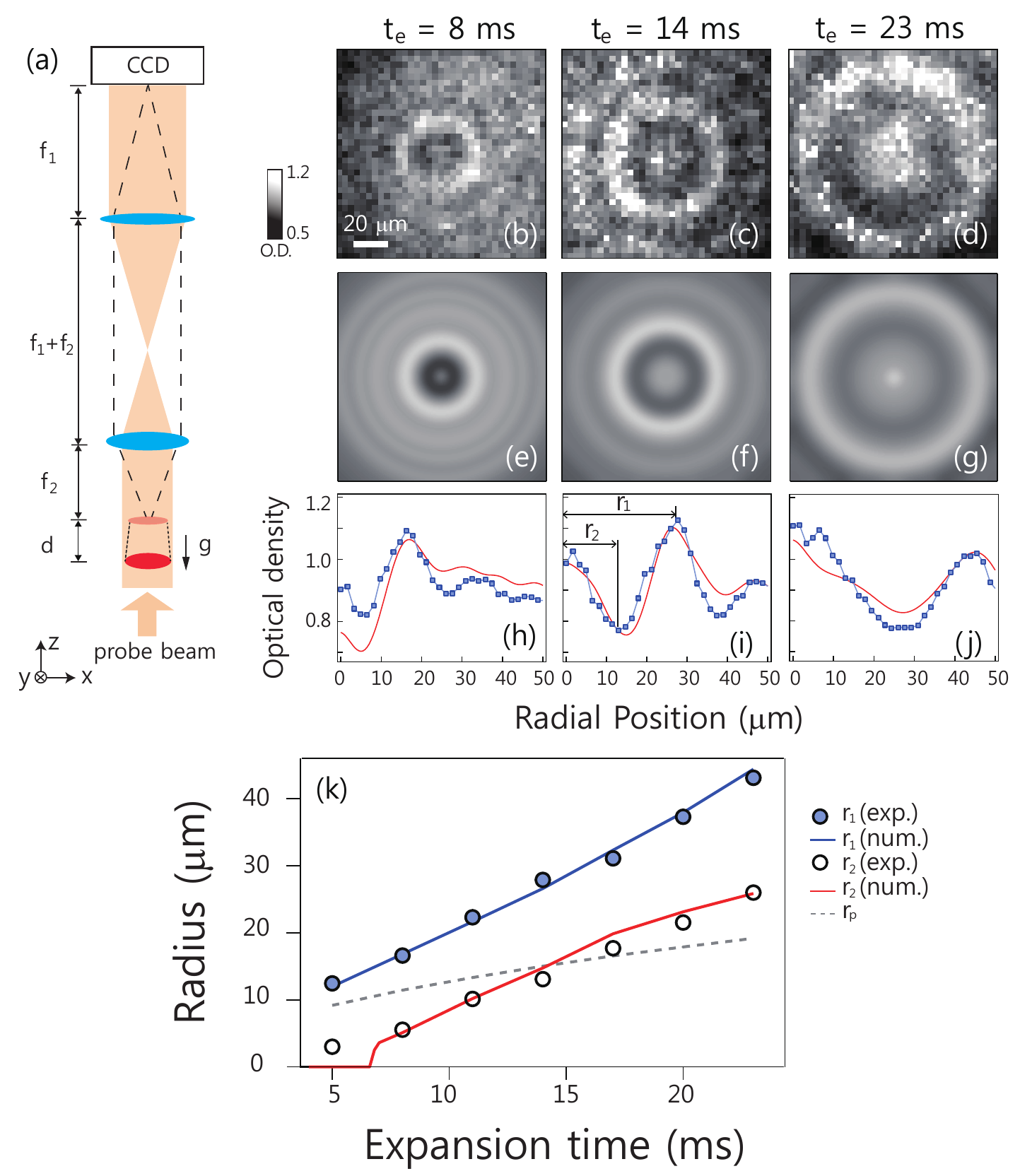}
\caption{
(Color online) (a) Schematic diagram of the imaging system. The condensate falls under gravity during expansion and the defocus distance $d=\frac{1}{2}gt_e^2$. Defocused images of a $l$=1 vortex core for various expansion times $t_e$: (b)-(d) experimental data and (e)-(g) numerical calculations. (h)-(j) The corresponding radial profiles from the experimental data [(b)-(d), blue square] and the numerically constructed defocused images [(e)-(g), red line]. (k) Radius $r_1$ ($r_2$) of the first bright (dark) ring versus expansion time. The dashed line is the radius $r_p$ for $l=1$ [Fig.~1(a)].}
\label{fig3}
\end{figure}

\section{Defocus effect}

In our previous experiment~\cite{PhaseFluc}, similar ring-shaped density ripples were observed in a mechanically perturbed and freely expanding quasi-2D condensate. Based on the observation of their apparent ring symmetry and long lifetime, we interpreted them as vortex excitations induced by the mechanical perturbation. However, the core region was filled up after a certain expansion time and this is unexpected in vortex expansion dynamics because of the phase singularity at the vortex core. We find that it was caused by defocussing due to the free fall of the condensate during the expansion time. In this section, we present an analysis of the defocused images of the vortex cores.

Figure 3(a) presents a schematic of the 4$f$ imaging system used in our experiments. When the condensate is released from the optical trap, it starts falling due to the gravity and its distance to the object plane of the imaging system increases as $d=\frac{1}{2}g t_e^2$, where $g$ is the gravitational acceleration. Then, the image recorded on the CCD camera is the near-field diffraction pattern of the probe beam after propagating by distance $d$ from the condensate. The depth of field in the imaging system is about 20~$\mu$m, so the defocus effect due to the free fall becomes significant already at $t_e>2$~ms. The in-focus images in Fig.~\ref{fig1} were taken by adjusting the axial position of the CCD camera to have a smallest vortex core for a given expansion time.

We numerically construct the defocused image from the calculated density distribution $n(r,t_e)$. Ignoring the sample thickness $\sim [\frac{\hbar}{m \omega_z}(1+\omega_z^2 t_e^2)]^{1/2}<60~\mu$m for $t_e<23$~ms, we assume the probe beam after the sample to have a uniform phase and an intensity distribution of $I(r)=I_0\exp[-\sigma n(r)]$, where we set the central optical density $\sigma n_0 = 1.1$ as measured in experiment. From the Fresnel diffraction theory~\cite{Mingu}, the beam intensity distribution $I_d(r)$ after propagating by distance $d$ is given as
\begin{equation}\label{eq3}
 I_d{(r)}=\frac{4\pi^2}{\lambda^2 d^2}\Big | \int_{0}^{\infty} \sqrt{I(r_1)} J_0(\frac{2\pi r_1r}{\lambda d})\exp(-i\frac{\pi r^2_1}{\lambda d})r_1 dr_1 \Big|^2,
\end{equation}
where $\lambda=589$~nm is the wavelength of the probe beam. Finally, we convolute $I_d(r)$ with a Gaussian function with $1/e^2$ width of 5~$\mu$m to take into account the finite imaging resolution.

The constructed defocused images reveal the core-filling behavior consistent with the experimental observation [Fig.~3(e)-(g)]. Furthermore, the radial profiles show good quantitative agreement with the experiment data over the whole range of the expansion time $5~\textrm{ms}<t_e<25~\textrm{ms}$ [Fig.~3(h)-(j)]. When we modify the density distribution $n(r)$ deliberately, e.g., to have no ripples but a constant value for $r>r_p$, the resultant defocused images are significantly deviated from the experiment data. This shows that the defocused image is sensitive to the details of the density distribution. Controlled defocussing might be used to probe small features below the spatial resolution of the imaging system.

It is well known in the diffraction theory that when a light propagates through an aperture of size $a$, its diffraction pattern after the aperture is determined by the Fresnel number, $\pi a^2/\lambda d $, where $d$ is the propagation distance from the aperture. Since the shape of the concentric ripples around the vortex core is almost preserved during expansion, we expect the core-filling behavior to occur when the normalized propagation distance $D=\lambda d / \pi r_p^2>D_c$. With $r_p\propto \sqrt{t_e}$ and $d=\frac{1}{2}g t_e^2$, $D\propto t_e$, suggesting that the core-filling always occurs in the defocused image after a certain expansion time (Fig.~4). We numerically find that the core region begins to have a local maximum at $t_e\approx 6.8, 9, 11$~ms for $l=1, 2, 3$, respectively, giving the critical value $D_c\approx0.36$.

\begin{figure}
\includegraphics[width=6.0cm]{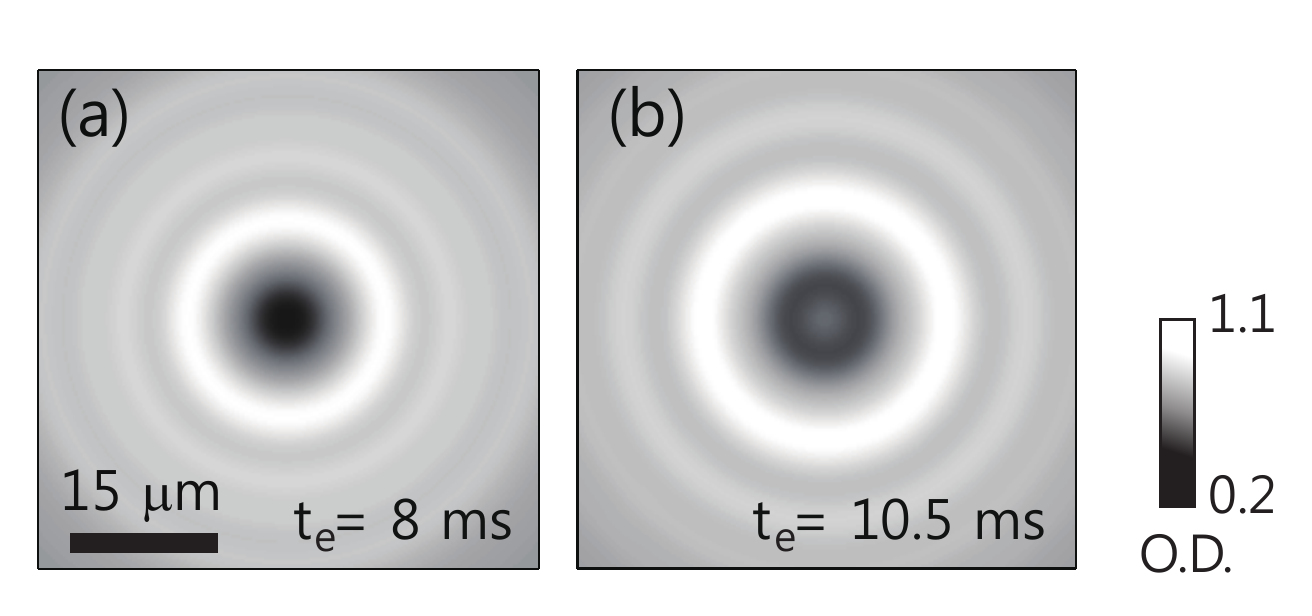}
\caption{
Defocused images for a doubly-charged ($l$=2) for (a) $t_e=$8~ms and (b) 10.5~ms. The solid circles in Fig.~2(b) indicate the corresponding positions in the $r_p$-$t_e$ plane, where the point for (b) is below the dotted line so that the core region in the defocused image shows a local maximum.}
\label{fig3}
\end{figure}

\section{Conclusion and Outlook}

We have studied the evolution of the quantized vortex core in a freely expanding quasi-2D BEC. The quantitative agreement between the experiment data and the numerical results demonstrates that the free expansion of the quasi-2D condensate is well described as non-interacting ballistic expansion. We have investigated the defocus effect caused by the free fall of the condensate under gravity, and provided a quantitative explanation on the ring-shaped density ripples observed in our previous experiment~\cite{PhaseFluc}.

The defocus effect can affect the measurement of the power spectrum of the density fluctuations in a freely expanding 2D Bose gas~\cite{private}. Recently, we performed in-focus measurements of the power spectrum and experimentally verified that the spectral peak positions are consistent with the theoretical prediction in Ref.~\cite{Igor}.

One interesting extension of this work is to study free expansion of a quasi-2D BEC containing a vortex-anti-vortex pair and investigate how large the pair should be to be detected by free expansion. Vortex-pair excitations are important to understand 2D superfluidity~\cite{BKT}. Thermally activated vortex pairs prevail in a 2D superfluid and the superfluid-to-normal phase transition occurs when thermal breaking of the pairs into free vortices becomes energetically preferred. Recently, we have succeeded in detecting thermal vortex pairs by applying a radial compression to enhance the visibility of vortex cores~\cite{BKTpaper}. However, it is still desirable to develop an experimental method to detect thermal vortices in a 2D Bose gas in a less perturbative manner.

\begin{acknowledgments}
We thank Woo-jin Kim for assistance in numerical simulation, and Igor Mazets and Tim Langen for helpful discussions. This work was supported by the NRF of Korea (2011-0017527, 2008-0062257, 2013-H1A8A1003984) and the POSCO TJ Park Foundation.
\end{acknowledgments}

\end{document}